\documentclass{aa} 
\usepackage[authoryear]{natbib}
\bibpunct{(}{)}{;}{a}{}{,}
\usepackage{amsmath}
\usepackage{graphicx}
\usepackage{txfonts}
\usepackage{lscape}

\begin{document}

\titlerunning{Improving SPH using {\sl IAD$_0$}} 
\authorrunning{Garc\'{i}a-Senz, Cabez\'on and Escart\'{i}n}

\title{Improving smoothed particle hydrodynamics with an integral approach to calculating gradients}

\author{Domingo Garc\'{i}a-Senz
        \inst{1},
       Rub\'en M. Cabez\'on \inst{2},
       \and
        Jos\'e Antonio Escart\'{i}n\inst{1}
       }
\institute
{Dept. de F\'{i}sica i Enginyeria Nuclear, Universitat Polit\`ecnica de Catalunya. Compte d'Urgell 187, 08036 Barcelona (Spain) and Institut d'Estudis Espacials de Catalunya.  Gran Capit\` a 2-4, 08034 
 Barcelona, Spain \\
 \and
Departement Physik. Universit\"at Basel. Klingelbergstrasse 82, 4056 Basel, Switzerland and Dept. de F\'{i}sica i Enginyeria Nuclear, Universitat Polit\`ecnica de Catalunya. Compte d'Urgell 187, 08036, 
 Barcelona, Spain\\}

\date{}

\abstract
{The smoothed particle hydrodynamics (SPH) technique is a well-known numerical method that has been applied to simulate the evolution of a wide variety of systems. Modern astrophysical applications of the method rely on the Lagrangian formulation of fluid Euler equations which is fully conservative. A different scheme, based on a matrix approach to the SPH equations is currently being used in computational fluid dynamics (CDF). These matrix formulations achieve better  interpolations of the physical magnitudes but they are, in general, not fully conservative. The matrix approach to the Euler equations has never been used in astrophysics.}  
{We develop and test a fully conservative SPH scheme based on a tensor formulation that can be applied to simulate astrophysical systems.} 
{In the proposed scheme, derivatives are calculated from an integral expression that leads to a tensor (instead of a vectorial) estimation of gradients and reduces to the standard formulation in the continuum limit. The new formulation improves the interpolation of physical magnitudes, leading to a set of conservative equations that resembles those of standard SPH. The resulting scheme is verified using a variety of well-known tests, all of them simulated in two dimensions. We also discuss an application of the proposed tensor method to astrophysics by simulating the stability of a Sun-like polytrope calculated in three dimensions.}    
{The proposed scheme is able to improve the results of standard SPH in the two-dimensional tests, especially in the simulation of subsonic hydrodynamic instabilities. Our results for the stability of the Sun-like polytrope suggest that the new method can be used in astrophysics to carry out three-dimensional calculations with a computational cost that is only slightly higher (i.e. $\le 50\%$ for a serial code) than that of a standard SPH formulation.} 
{A formalism based on a matrix approach to Euler SPH equations was developed and checked. The new scheme is more accurate because of the re-normalization imposed on the interpolations, which is fully conservative and probably less prone to undergo the tensile instability. The analysis of several test cases suggest that the method may improve the simulation of both subsonic and supersonic systems. An application of the tensor method to astrophysics is, for the first time, successfully carried out. These encouraging results indicates that more work should be invested in the applications of matrix SPH formulations to astrophysics.}   


\maketitle

\section{Introduction}

The hydrodynamic method known as smoothed particle hydrodynamics (SPH) is a grid-less Lagrangian approach to continuum mechanics devised by \citet{gm77}~and \citet{lucy77}. As demonstrated in an endless amount of applications to computational fluid dynamics, ranging from large-scale astrophysical simulations to nuclear physics on the Fermi level, SPH is a robust and confident way to simulate the dynamical evolution of deformed systems. The success of SPH relies strongly on the way gradients are estimated. For example, the value of density in a given spatial coordinate is obtained from particles located in the neighborhood using an interpolating function called {\sl kernel}. As the kernel is an analytical differentiable function, it is easy to obtain gradients just evaluating the gradient of this weighting kernel function. Then it is quite straightforward to write the Euler equations of fluid mechanics in terms of the kernel and its derivatives \citep{monaghan92,monaghan05}.

Despite the success, SPH has also a number of shortcomings and weak points that make the technique less useful than it should be to handle a number of well-identified problems. First of all, the numerical noise is usually larger than in other techniques, which poses a problem for applications involving very subsonic movements, as in the case of hydrodynamical instabilities. Another difficulty has to do with the use of the artificial viscosity (AV) formalism to handle shock waves and smooth the inherent presence of noise. Several solutions and recipes have been invoked to handle these problems, none of them totally satisfactory, among them: variable AV coefficients \citep{morris97}, artificial heat fluxes to smooth pressure discontinuities \citep{price08}, non-standard approximations to the momentum equation \citep{abel11}, and a refined treatment of kernel interpolations as in the technique known as moving least squares interpolation \citep[i.e. {\sl MLS};][]{dilts99}.

In this paper we present an approach to the momentum and energy equations that combines a novel way to estimating gradients and the variational principle, and allows an improved modeling of physical phenomena, especially in systems subjected to small perturbations. The proposed formalism, based on an integral approach to the derivatives ({\sl IAD}), includes the standard formulation as a particular case. It leads to a matrix formulation similar to that of MLS methods, but the form of the SPH equations in {\sl IAD} remains closer to that of standard SPH. The text is organized as follows. In Sect. \ref{section2} the mathematical formalism for calculating gradients is described and discussed. In Sect. \ref{section3} we develop a formulation of the momentum equation compatible with the variational Euler-Lagrange principle in the SPH framework, and propose a consistent and fully conservative energy equation. Section \ref{section4} is devoted to validating the hydrocode through the simulation of four known tests where standard SPH has, to a smaller or larger degree, difficulties. In Sect. \ref{section5} we elaborate an extension of the method that by combining the exact estimation of the derivative of linear functions and good momentum and energy conservation, can be applied to simulate linearized systems. The ability of the proposed method to simulate the structure of real three-dimensional (3D) astronomical bodies is discussed in Sect. \ref{section6}. Finally, the main conclusions of our work, as well as some comments about the shortcomings of the developed scheme and future lines of improvement are outlined in the last section, which is devoted to conclusions.

\section{Integral approach to first derivatives}
\label{section2}

There is an ample literature describing the treatment of the first and second derivatives in SPH \citep{monaghan05, rosswog09}. In the following we take a different route to estimating first derivatives which leads to a more general expression than the most common extant derivative procedures. Our method resembles to those known as {\sl MLS} methods where interpolations relative to gradient estimation were constrained to enhance linearity by imposing a minimization procedure on the errors.

Inspired by the treatment of second derivatives \citep{brookshaw85} we define an integral expression from which we will estimate gradients. As in the case of second derivatives, it is expected that an integral approach will lead to a less noisy estimation of first derivatives. Therefore, we define

\begin{equation}
I({\bf {r}})= \int_{V} \left[f({\bf {r'}})-f({\bf {r}})\right] ({\bf {r'}}-{\bf {r}}) W(\vert {\bf {r}'}-{\bf {r}}\vert, h) dr'{^3}\,,
\label{int}
\end{equation}

\noindent where $f(r)$ is any differentiable function and $W(\vert {\bf {r}'}-{\bf {r}}\vert, h)$ is a spherically symmetric interpolating kernel, usually a sharp-like Gaussian of width $h$ (where $h$ is usually referred as the smoothing length). The size of $h$ is often interpreted as the local resolution of the simulation. Expanding the factor $\left[f({\bf {r'}})-f({\bf {r}})\right]$ to first order

\begin{equation}
f({\bf {r'}})-f({\bf {r}})=\bf\nabla f\cdot({\bf {r'}}-{\bf {r}}) + \mathrm {higher~ order~ terms}
\end{equation}

\noindent and writing

\begin{equation}
({\bf {r'}}-{\bf {r}})= ({x'_1}-x_1){\bf i}+(x'_2-x_2){\bf j}+(x'_3-x_3){\bf k}\,,
\end{equation}

\noindent
equation (\ref{int}) can be expressed as a matrix equation:

\begin{equation}
\left[
\begin{array}{c}
\partial f/\partial x_1\\
\partial f/\partial x_2\\
\partial f/\partial x_3\\
\end{array}
\right]
=
\left[
\begin{array}{ccc}
\tau_{11} & \tau_{12} & \tau_{13} \\
   \tau_{21}&\tau_{22}&\tau_{23} \\
   \tau_{31}&\tau_{32}&\tau_{33}
\end{array}
\right]^{-1}
\left[
\begin{array}{c}
I_1\\
I_2\\
I_3\\
\end{array}
\right]\,,
\label{matrix}
\end{equation}

\noindent where

\begin{equation}
\tau_{ij}=\int (x'_i-x_i)(x'_j-x_j) W(\vert {\bf {r}'}-{\bf {r}}\vert, h) dr'{^3}\,;i,j=1,3
\label{tauij}
\end{equation}

\noindent and

\begin{equation}
I_k=\int \left[f({\bf {r'}})-f({\bf {r}})\right] ({x_k'}-{x_k}) W(\vert {\bf {r}'}-{\bf {r}}\vert, h) dr'{^3}\,;k=1,3\,.
\end{equation}

Once the linear system in Eq. (\ref{matrix}) is solved we get the value of the gradient of the function $f$. In the case of spherically symmetric kernels, $\tau_{11}=\tau_{22}=\tau_{33}$ and $\tau_{ij}=\tau_{ji}=0$, $i \ne j$. Taking for instance the Gaussian kernel

\begin{equation}
W^G(u,h)=\frac{1}{\pi^{n/2}}\exp(-u^2)\,,
\end{equation}

\noindent where $n$ is the dimensionality of space and $u=\frac{(\vert {\bf {r}'}-{\bf {r}}\vert)}{h}$, it leads to $\tau_{ii}=\frac{h^2}{2}$. Therefore, for the Gaussian kernel the gradient of $f$ reduces to

\begin{equation}
 {\bf\nabla f} = \int \left[f({\bf {r'}})-f({\bf {r}})\right] {\bf\nabla W}(\vert {\bf {r}'}-{\bf {r}}\vert, h) dr'{^3}\,,
\label{nablaf}
\end{equation}

\noindent which is the same expression as Eq. (2.13) of \citet{monaghan05} even though it has been obtained through a different procedure. Since the standard kernels are spherically symmetric Eq. (\ref{nablaf}) can be simplified to

\begin{equation}
{\bf\nabla f} = \int f({\bf {r'}}) {\bf\nabla W}(\vert {\bf {r}'}-{\bf {r}}\vert, h) dr'{^3}\,,
\end{equation}

\noindent which in SPH parlance becomes

\begin{equation}
{\bf\nabla_a f}= \sum_b \frac{m_b}{\rho_b}f_b {\bf\nabla W}_{ab}(h_a)\,,
\label{nablafSPH}
\end{equation}

\noindent which because of its simplicity and good performance has become the most popular procedure for calculating first derivatives in SPH. Therefore, Eq. (\ref{nablafSPH}) is a {\sl particular case} of {\sl IAD}, Eq. (\ref{matrix}).

Nevertheless, many of the above symmetry properties of tensor ${\bf\mathcal{T}}\equiv \{\tau_{ij}\}$ are lost when integrals are converted into finite summations. For example, the elements on the diagonal of matrix ${\bf\mathcal{T}}$ in Eq. (\ref{matrix}) do not necessarily have the same value and elements outside the diagonal can differ from zero. However, the formulation of the first derivative in terms of the matrix Eq. (\ref{matrix}) has a very interesting feature: the derivative of a linear function is always exact by construction. The demonstration of such property is straightforward in 1D SPH:

\begin{equation}
\left(\frac{df}{dx}\right)_a=\frac{\sum_b \frac{m_b}{\rho_b}(f_b-f_a)(x_b-x_a)W_{ab}}{\sum_b \frac{m_b}{\rho_b}(x_b-x_a)^2 W_{ab}}\,,
\end{equation}

\noindent taking $f_a=px_a+q, f_b=px_b+q$ the expression above gives $(\frac{df}{dx})_a=p$.

In {\sl two dimensions}, the explicit solution for Eq. (\ref{matrix}) is written as

\begin{equation}
\frac{\partial{f}}{\partial x_1} = c_{11} I_1 + c_{12} I_2\,;\qquad\frac{\partial{f}}{\partial x_2} = c_{21} I_1 + c_{22} I_2\,,
\end{equation}

\noindent where

\begin{equation}
\label{cij}
\begin{array}{ll}
c_{11} =\left(\tau_{11}-\frac{\tau_{12}^2}{\tau_{22}}\right)^{-1}\,;\qquad& c_{12}=-\left(\frac{\tau_{12}}{\tau_{22}}\right)\left(\tau_{11}-\frac{\tau_{12}^2}{\tau_{22}}\right)^{-1}\,;\\
c_{21} =-\left(\frac{\tau_{12}}{\tau_{11}}\right)\left(\tau_{22}-\frac{\tau_{12}^2}{\tau_{11}}\right)^{-1}\,;& c_{22}=\left(\tau_{22}-\frac{\tau_{12}^2}{\tau_{11}}\right)^{-1}\,.
\end{array}
\end{equation}

Unfortunately, direct application of the tensor scheme above leads to SPH equations of movement that are not fully conservative. Nevertheless, a slight modification of the scheme makes it possible to build exact conservative equations by taking advantage of the symmetry properties of the kernel to discretize the vector {\bf I}(r) as

\begin{equation}
\begin{split}
I({\bf {r}})=\int \left[f({\bf {r'}})-f({\bf {r}})\right] ({\bf {r'}}-{\bf {r}}) W(\vert {\bf {r}'}-{\bf {r}}\vert, h) dr'{^3} \simeq\\
\sum_b\frac{m_b}{\rho_b} f({\bf r_b})({\bf {r_b}}-{\bf {r_a}}) W(\vert {\bf {r_b}}-{\bf {r_a}}\vert, h_a)\,.
\end{split}
\label{approxI}
\end{equation}

Thereafter, the formulation that results after substituting Eq. (\ref{approxI}) into Eq. (\ref{matrix}) is labeled {\sl IAD$_0$}. This restricted interpretation of Eq. (\ref{matrix}) implies that exact evaluation of the derivative of linear functions cannot be achieved if we are to also preserve the exact conservation of momentum and energy. Nevertheless, we present below strong indications that using {\sl IAD$_0$}~leads to a better evaluation of the derivative of linear functions than the standard procedure given by Eq. (\ref {nablafSPH}), especially when a small or moderate number of neighbors is used to compute summations. We can illustrate this point in the following numerical experiments.

As a first test, we simulate a static system where we compare the gradient of a linear distribution of density in one dimension obtained with {\sl IAD}, {\sl IAD$_0$} and standard schemes as a function of the number of neighbors $n_b$. A linear density  profile $\rho(x)=1+x$~was obtained aligning N=100 equidistant particles of adequate mass along the $x$-axis. The density was calculated using

\begin{equation}
\rho_a=\sum_b m_b W_{ab}(\vert {\bf r_b-r_a\vert},h_a)\,.
\label{density}
\end{equation}

Note that even though {\sl IAD} should provide the precise derivative, in practice it never does owing to the small errors in the calculation of density. It is also worth noting that the standard calculation of the derivative has two additional potential sources of error. For a linear density profile $\rho_b=\rho_a+p (x_b-x_a)$, and using Eq.(\ref{nablafSPH}), we have

\begin{equation}
\begin{split}
\left(\frac{d\rho}{dx}\right)_a^{STD}&=\sum_b \frac{m_b}{\rho_b}\rho_a\nabla_{a} W_{ab}+p\sum_b \frac{m_b}{\rho_b}(x_b-x_a)\nabla_a W_{ab}\\
&=\epsilon_1+ \epsilon_2~p\,,
\label{gradlinealstd}
\end{split}
\end{equation}

\noindent where $p$~is the 'exact' derivative and $\epsilon_1\simeq 0$,~$\epsilon_2\simeq 1$. On the other hand, there is only one source of error, $\epsilon_3\simeq 0$,~ if {\sl IAD$_0$} is used, given by

\begin{equation}
\left(\frac{d\rho}{dx}\right)_a^{IAD_0}=\frac{\sum_b \frac{m_b}{\rho_b}\rho_a(x_b-x_a)W_{ab}}{\sum_b \frac{m_b}{\rho_b}(x_b-x_a)^2 W_{ab}}+ p= \epsilon_3 + p\,.
\label{gradlinealIAD0}
\end{equation}

In Fig. \ref{figure1} (upper and bottom left) we represent the value of the relative error in the derivative $\epsilon=\vert \frac{d\rho}{dx}-p\vert/p$, with respect to the analytical value $p$ as a function of the smoothing length $h$, normalized to the inter-particle distance $\Delta$, for the Gaussian and cubic spline kernels respectively. Independently of the kernel, the error when full {\sl IAD} linear interpolation was used is always much smaller than for the other methods. When {\sl IAD$_0$} or the standard derivative were used, the error increased appreciably. The error is large when the smoothing length is shorter than the inter-particle distance but decreases rapidly when $h$ increases, as expected. Nevertheless, the error in the standard calculation always remained larger, especially for a small or moderate number of neighbors ($n_b\le 5$) regardless of the kernel we used, but especially for the cubic spline. Since the particle sample is highly ordered, the error curve of the Gaussian kernel is much smoother than that of the cubic spline owing to the infinite range of the kernel. We verified that when the  Gaussian is truncated to $2h$, the error profile becomes qualitatively similar to that of the cubic spline. However, the error profile for {\sl IAD$_0$}~follows almost exactly the error curve for the gradient of density calculated as a simple quotient $\frac{\rho_b-\rho_a}{x_b-x_a}$ for adjacent particles. In this sense, {\sl IAD$_0$} seems to be more coherent with the computed density distribution of the sample than {\sl IAD}, a trend that also holds in 2D, as shown below.
       
The contributions to the total error of $\epsilon_1, \epsilon_2$~and $\epsilon_3$ in Eqs. (\ref{gradlinealstd}) and (\ref{gradlinealIAD0}) are shown in the rightmost part of Fig. 1. The contribution of $\epsilon_1$~and $\epsilon_3$ are similar but the error $\epsilon_2$, which affects only the standard scheme and exhibits a large oscillation in the case of compact supported interpolators, represents the main channel to the total error in the derivative. In general, tensor methods become more reliable for higher dimensionality. Thus, we conducted a similar numerical experiment in more than one dimension where the matrix approach can potentially be more beneficial.

As a second test, a bi-dimensional system was set using a sample of N=62\,500 equally spaced particles on a two-dimensional lattice. The mass of the particles along the {\it x}-axis was conveniently modified to reproduce a linear density profile, $\rho(x,y)=1+x$. The first derivative of the density, $\frac{\partial\rho}{\partial x}$~for different number of neighbors was obtained using Eqs. (\ref{matrix}) and (\ref{nablafSPH}). In the upper row of Fig. \ref{figure2} we show the relative error in the derivative,  $\epsilon=\vert \frac{\partial\rho}{\partial x}-p\vert/p$, with respect to the analytical value $p$~ as a function of the smoothing length. Calculations were carried out using the Gaussian and the cubic spline kernels. The trend here is similar to that of the 1D calculation but magnitudes fluctuate less, especially in the case of the cubic spline. When the number of neighbors is large the error is small in all three methods, but tensor schemes are more accurate than the standard procedure. Nevertheless, for shortening smoothing lengths, h $\le 1.5\Delta$, the error curve steepens and for h $< \Delta$~standard SPH begins to give unacceptable results while the tensor estimations of the derivative remain applicable. This result also agrees with the 1D calculation shown in Fig. 1, which shows that working with $n_b\le 3, 4$ in 1D may give an appreciable error in the derivative of a linear function if Eq. (\ref{nablafSPH}) is used to compute the derivatives. As in the 1D case, the gradient calculated as a direct fraction $\frac{\rho_b-\rho_a}{x_b-x_a}$ (crosses in Fig. 2) remained closer to the results obtained using {\sl IAD$_0$}. 

Finally, we considered a spherically symmetric linear profile $\rho(x,y)=1+r$, where $r=\sqrt{x^2+y^2}$. In the bottom row of Fig. \ref{figure2}, the value of the relative error $\epsilon$ is shown as a function of $h(\Delta)$. The profile of the error follows a slightly different trend than in the previous case. As before, the error is large when the smoothing length is shorter than the inter-particle distance but in the case of the Gaussian kernel there is a minimum in the error profile at $h\simeq 1.1\Delta$. From that point on the error smoothly increases. This behavior is not so clear in the case of the cubic spline where the error curve roughly stabilizes above $h=2\Delta$. There is, therefore, an optimal value of $h$ in this case that minimizes the errors in the first derivative. Close to the center linearity is lost because of the symmetry imposed on the distribution. When the value of $h$~is large, interpolations are affected by that geometrical constraint and, at some point, the advantages of working using a large smoothing length are lost. Nevertheless, the errors calculated using {\sl IAD$_0$} are always smaller than that of the standard scheme. Again the gradient of density estimated as a simply quotient $\frac{\rho_b-\rho_a}{r_b-r_a}$ is more closely reproduced by {\sl IAD$_0$} than by the other schemes.

\section{The momentum and energy SPH equations using IAD$_{\mathbf 0}$}
\label{section3}

Nowadays the most accurate formulation of the momentum equation in SPH comes from the variational principle. It has been shown \citep[and references therein]{monaghan05} that the solution of the Euler-Lagrange equations leads naturally to a very symmetric scheme including the effects of spatial gradients on the smoothing length. The resulting equation conserves momentum by construction and, equally important, ensures perfect energy conservation for non-dissipative systems. It is shown below how the tensor approach built using the {\sl IAD$_0$} scheme can also be compatible with the variational principle. 

The Lagrangian of the system is evaluated as

\begin{equation}
L=\sum_b m_b \left(\frac{1}{2} v_b^2-u_b(\rho_b,s_b)\right)\,,
\label{lagrangian}
\end{equation}

\noindent where $v_b$, $u_b$ and $s_b$ are the velocity, specific internal energy and specific entropy of particle $b$. Using this Lagrangian and assuming isentropic evolution, the Euler-Lagrange equations subjected to the constraint $\rho h^3=\mathrm {constant}$, lead to the equation of movement \citep{springel02}\footnote{In that paper the $\nabla h$ contribution is estimated using the Lagrange multipliers technique.}

\begin{equation}
m_a \ddot {x_a}=-\sum_b m_b\frac{P_b}{\rho_b^2\Omega_b}\nabla_a\rho_b\,,
\label{lagr1}
\end{equation}

\noindent where $\Omega_b=(1-\partial\rho/\partial h~ \sum_c m_c~ \partial W_{bc}/\partial h)$~is a term accounting for the gradient of $h$. The $i$-component  in Eq. (\ref{lagr1}) can be written

\begin{equation}
m_a\ddot x_{i,a}=-\sum_{b\neq a} m_b\frac{P_b}{\rho_b^2\Omega_b}\frac{\partial \rho_b}{\partial x_{i,a}}-m_a \frac{P_a}{\rho_a^2\Omega_a}\frac{\partial\rho_a}{\partial x_{i,a}}\,.
\label{lagr2}
\end{equation}
\noindent
An estimation of the density gradients using the tensor Eq. (\ref{matrix}) is

\begin{equation}
{\bf\nabla}\rho={\bf\mathcal{C}} {\bf I}\,,
\label{acceleration}
\end{equation}

\noindent
where ${\bf \mathcal{C}=\mathcal{T}}^{-1}$. Elements of matrix $\mathcal{T}$ are those defined in Eq. (\ref{tauij}) after changing integrals to summations. For particle $a$, they are defined to be

\begin{equation}
\tau_{ij,a}=\sum_b \frac{m_b}{\rho_b}(x_{i,b}-x_{i,a})(x_{j,b}-x_{j,a})W_{ab}(h_a)\,,
\label{tauijsph}
\end{equation}

\noindent and the $j$-component of vector ${\bf I}$ in the {\sl IAD$_0$} approach is

\begin{equation}
I_{j,a}=\sum_b~m_b
(x_{j,b}-x_{j,a}) W_{ab}(h_a)\,.
\label{Iksph}
\end{equation}
\noindent
Thus, for particle $a$~the $i$-component of density gradient is

\begin{equation}
\begin{split}
\frac{\partial\rho_a}{\partial x_{i,a}}=&\sum_{j=1}^d c_{ij,a}(h_a)~I_{j,a}(h_a)=\\
&\sum_{j=1}^d\sum_b^{n_b} m_b c_{ij,a}(h_a) (x_{j,b}-x_{j,a})~W_{ab}(h_a)\,,
\label{lagr3}
\end{split}
\end{equation} 

\noindent where $ c_{ij,a}$ are the elements of matrix $\mathcal{C}$ associated with particle $a$, and $d$ is the dimension of the space. Explicit expressions for these elements in cartesian 2D were given by Eq. (\ref{cij}). Equation (\ref{lagr3}) can be used to directly compute the second term on the right of Eq. (\ref {lagr2}). 

We can evaluate, in the same way, for particle $b$

\begin{equation}
\frac{\partial\rho_b}{\partial x_{i,b}}=
\sum_{j=1}^d\sum_c^{n_b} m_c c_{ij,b}(h_b) (x_{j,c}-x_{j,b})~W_{bc}(h_b)\,.
\label{lagr3b}
\end{equation} 

To calculate $\frac{\partial \rho_b}{\partial x_{i,a}}$ we use Eq. (\ref {density}):
 
\begin{equation}
\frac{\partial\rho_b}{\partial x_{i,a}}= -m_a\frac{\partial W_{ab}(h_b)}{\partial x_{i,b}}\,.
\label{lagr4}
\end{equation} 

\noindent The last term in Eq. (\ref{lagr4}) also appears during the evaluation of $\nabla_b\rho_b$, namely

\begin{equation}
\frac{\partial\rho_b}{\partial x_{i,b}}=\sum_c m_c \frac{\partial W_{bc}(h_b)}{\partial x_{i,b}}= ... + m_a \frac{\partial W_{ab}(h_b)}{\partial x_{i,b}} + ...
\label{lagr5}
\end{equation}

\noindent Comparing Eqs. (\ref{lagr5}), (\ref{lagr4}) and (\ref{lagr3b}), we finally get

\begin{equation}
\frac{\partial\rho_b}{\partial x_{i,a}}= \sum_{j=1}^d m_a c_{ij,b}(h_b)(x_{j,b}-x_{j,a})W_{ab}(h_b)\,.
\label{lagr6}
\end{equation}

\noindent Substituting Eqs. (\ref{lagr3}) and (\ref{lagr6}) into Eq. (\ref{lagr2}), the $i$-component of the momentum equation for particle $a$ is given by 

\begin{equation}
\ddot x_{i,a}=-\sum_{b=1}^{n_b} m_b\left(\frac{P_a}{\Omega_a\rho_a^2}\mathcal A_{i,ab}(h_a)+\frac{P_b}{\Omega_b\rho_b^2}\mathcal A'_{i,ab}(h_b)\right)\,,
\label{momentumL}
\end{equation}

\noindent where 

\begin{align}
{\mathcal A}_{i,ab}(h_a) &=\sum_{j=1}^{d} c_{ij,a}(h_a) (x_{j,b}-x_{j,a}) W_{ab}(h_a)\,,\label{lagr7}\\
{\mathcal A}'_{i,ab}(h_b) &=\sum_{j=1}^{d} c_{ij,b}(h_b) (x_{j,b}-x_{j,a}) W_{ab}(h_b)\,.\label{lagr8}
\end{align}

\noindent It is then straightforward to derive the appropriate energy equation

\begin{equation}
\left(\frac{du}{dt}\right)_a=\frac{P_a}{\Omega_a~\rho_a^2}\sum_{b=1}^{n_b}\sum_{i=1}^d m_b (v_{i,a}-v_{i,b})\mathcal A_{i,ab}(h_a)\,.
\label{energy}
\end{equation}

\noindent Since the magnitudes $\mathcal A_{i,ab}$, defined by Eqs. (\ref{lagr7}) and (\ref{lagr8}), are antisymmetric, the conservation properties of the {\sl IAD$_0$} formulation are identical to those of standard SPH.

As in standard SPH, an artificial viscosity term has to be added to stabilize the numerical scheme and handle shocks. The AV increases the effective pressure in those zones of the fluid that undergo compression. Including the viscous acceleration, the momentum equation is given by 

\begin{equation}
\ddot x_{i,a}=-\sum_{b=1}^{nb} m_b\left(\frac{P_a}{\Omega_a\rho_a^2}\mathcal A_{i,ab}(h_a)+\frac{P_b}{\Omega_b\rho_b^2}\mathcal A'_{i,ab}(h_b)+\Pi_{ab}~\widetilde{\mathcal A}_{i,ab}\right)\,,
\label{momentumLqij}
\end{equation}

\noindent where $\Pi_{ab}$ accounts for the viscous pressure


\begin{equation}
\Pi_{ab}=
\left\{\begin{array}{rclcc}
\frac{-\alpha c_{ab} \mu_{ab}+ \beta {\mu^2_{ab}}}{\bar\rho_{ab}} & \mathrm{for}~ & {\bf r}_{ab}\cdot {\bf v}_{ab} < 0, \\
 0~~ \mathrm{otherwise,}
\end{array}
\right.
\label{avis}
\end{equation}

\noindent where the symbols have their usual meaning and ${\bf r}_{ab}$ means ${\bf r}_a-~{\bf r}_b$. The coefficient $\mu_{ab}$ is

\begin{equation}
\mu_{ab}=\frac{\bar h_{ab}{\bf r}_{ab}\cdot {\bf v}_{ab}}{r_{ab}^2+0.01~\bar h^2_{ab}}\,.
\end{equation}

\noindent To preserve momentum conservation, we evaluate the arithmetic mean of $\mathcal A$

\begin{equation}
\widetilde{\mathcal A}_{i,ab}=\frac{1}{2}\left[{\mathcal A}_{i,ab}(h_a)+{\mathcal A'}_{i,ab}(h_b)\right]\,.
\label{atilde}
\end{equation}

The inclusion of artificial viscosity in the energy equation leads to 

\begin{equation}
\left(\frac{du}{dt}\right)_a=\sum_{b=1}^{n_b}\sum_{i=1}^d m_b(v_{i,a}-v_{i,b})\left(\frac{P_a}{\Omega_a~\rho_a^2}~\mathcal{A}_{i,ab}(h_a)+ \frac{\Pi_{ab}}{2}~\widetilde{\mathcal A}_{i,ab}\right)\,.
\label{energyqij}
\end{equation}

As a representative example of the ability of the new scheme to handle subsonic physics, we consider the stability of an inhomogeneous system in mechanical equilibrium. A sample of N=62\,500  particles were evenly distributed within a square lattice with periodic boundary conditions ensuring that the initial density was set to $\rho_0=1$~g.cm$^{-3}$. The density was then perturbed at random allowing maximum percentage variations of $5 \%$ across the lattice. The internal energy of each particle was adjusted to ensure isobaricity with $P_0=$ 1 dyn.cm$^{-2}$ leading to a sound speed value c$_s\simeq $~ 1.3 cm.s$^{-1}$. The evolution of the system was followed during almost a sound crossing time, t$_s\simeq 0.5$~s using the three schemes, {\sl IAD}, {\sl IAD$_0$} and standard SPH, with $n_b=30$ and 100. Calculations using full {\sl IAD} were carried out using the Eqs. (\ref{momentumLcomplete}) and (\ref{energycomplete}) described in Sect. \ref{section5}. Hereafter, the word {\sl standard}, STD,  refers to the modern, fully conservative, Lagrangian formulation of SPH that explicitly includes the gradient of the smoothing length parameter in the scheme (see for instance \citet{rosswog09} and references therein). The results of the simulations are summarized in Fig. \ref{figure3}, where we present the evolution of the standard deviation in the pressure $P(t)$ with respect to $P_0$ for the three aforementioned schemes. For this particular test it is clear that both tensor schemes provide a more persistent mechanical equilibrium than standard SPH during a sound crossing time, especially when full {\sl IAD} is used. As expected, the standard deviation in the pressure decreases as the number of neighbors increases.

\subsection{Computational issues}

The increase of the computational requirements of the integral method is more evident now. With respect to memory requirements, the nine coefficients (in 3D) of matrix ${\bf\mathcal{T}}$ and those belonging to matrix ${\bf \mathcal{C}}$ can share the same memory space. Thus, nine coefficients per particle have to be stored to compute $\mathcal{A}_{i,ab}$ in Eq. (\ref{momentumLqij}). While the increase in memory requirements are not a significant problem, the burden in the computational time is larger. First of all, one has to compute six of the nine coefficients of symmetric matrix ${\bf\mathcal{T}}$, Eq. (\ref{tauijsph}). Although these calculations involve simple operations, they have to be performed in a separate tree-walk because previous knowledge of density is necessary. Afterwards, matrix ${\bf\mathcal{T}}$ has to be inverted in order to calculate the coefficients $c_{ij,a}$ for particle $a$. Fortunately, these matrix inversions do not take much time because they can be completed out of the tree structure. Finally, one has to compute the nine quantities of coefficients $\mathcal{A}_{i,ab}$ associated with each pair of neighbors particles $a, b$; instead of the three that are necessary in the standard formulation, and use them to compute both the momentum and the energy equations.

It is obvious that the proposed method demands a larger computational effort than the standard SPH. Nevertheless, the extra burden could be small if the physical problem under simulation requires the calculation of long-range forces (i.e. gravity) or involves complex physics (i.e chemical or nuclear reactions, transport phenomena). In addition note that, according to Figs. \ref{figure1} and \ref{figure2}, for standard SPH to achieve a similar accuracy as {\sl IAD$_0$} in calculating the first derivative, the number of neighbors has to be larger. To maintain the spatial resolution, the total number of particles also has to be larger by the same factor. Therefore, changing both the number of neighbors and total number of particles could have a larger impact on the computational performance than using {\sl IAD$_0$}. 

\section{Hydrodynamic tests}
\label{section4}

The aim of this section is to check the theory described in Sects. \ref{section2} and \ref{section3} using four tests, all of them widely used in CFD. The first two simulations concern the growth of the KH and RT hydrodynamic instabilities, whose evolution always remains subsonic and, as we demonstrate, the fully conservative {\sl IAD$_0$} scheme leads to improved results with respect to the standard method for the same identical initial conditions. The last two tests are related to strong supersonic flows where shock waves take over, as in the wall heating shock and Sedov tests, where the use of the tensor route to estimate derivatives again provides more accurate results, especially for the wall heating shock test.

The simulations described in this section were carried out in a cartesian two dimensional scenario because of the higher achieved resolution, easiest initial setting and analysis of the results. Periodic boundary conditions were imposed in all tests except in the RT case, where particles from then top and the bottom of the box were fixed. A perfect gas equation of state (EOS) $P=(\gamma-1)\rho u$, where $u$ is the specific internal energy and $\gamma=5/3$, is always used. The coefficients $\alpha$, $\beta$ of artificial viscosity, in Eq. (\ref{avis}), were set to 1 and 2 respectively, except for the wall heating test where we took $\alpha=1.5,~\beta=3$. The cubic spline was used to perform the SPH interpolations but several calculations carried out with other interpolators (i.e. harmonic kernels with index n=3; \citet{cabezon08}) did not lead to significant differences. We used a self-adaptive smoothing length parameter which keeps constant the number of neighbors, $n_b$, of a given particle. Although the simulations described below were obtained for $n_b=100$, the same calculations carried out using $n_b=36$~neighbors did not provide significant differences. Corrections due to the gradient of the smoothing length were explicitly included in the {\sl IAD$_0$} formulation as it comes from the Lagrange-Euler variational principle.

In Table~1 we show a summary of energy and momentum conservation properties for the different calculated models. The tensor calculations are always more able to ensure momentum and energy conservation than the standard ones for the same elapsed time. The lack of energy conservation during the point-like Sedov explosion in all methods is due to the hard initial conditions. In contrast, the very good conservation of momentum in the supersonic numerical experiments is partially due to the spherical symmetry imposed to the initial configurations.

\subsection{Kelvin-Helmholtz instability}

The Kelvin-Helmholtz instability is a well-known test to check the ability of any hydrodynamic code to handle subsonic perturbations. This instability appears when there is a sufficient velocity shear in the interface layer between two fluids with different densities. Small perturbations of the velocity field in the orthogonal direction to the interface grow, leading to a mixing of both fluids. This is usually simulated in a box with periodic boundary conditions where two fluid regions are defined with densities $\rho_1$ and $\rho_2$ respectively. Both layers have opposite parallel velocities leading to a shear discontinuity in the contact interface. To develop the instability, a small perturbation is seeded in the interface as a sinusoidal mode of length scale $\lambda$.

In our case, we have simulated a central band of high density fluid ($\rho_1$) moving in a low-density medium ($\rho_2$) in a squared lattice of 1 cm side in the XY plane using N=62\,500 particles. The mass of the particles was arranged to obtain the correct density profile following a ramp function \citet{robertson10}. In this way, we smoothed the interface density jump to make it comparable to the SPH resolution using

\begin{equation}
f(y)=\frac{1}{A}~\frac{1}{1+\exp{\frac{2(y-0.25)}{\Delta y}}}~\frac{1}{1+\exp{\frac{2(0.75-y)}{\Delta y}}}\,,
\end{equation}

\noindent
where $A$ is a normalization constant and $\Delta y=0.05$~cm. Therefore, the density profile was given by

\begin{equation}
\rho(y)=\rho_2+(\rho_1-\rho_2)~f(y)\,,
\end{equation}

\noindent
 where $\rho_1=2$~g.cm$^{-3}$ and $\rho_2=1$~g.cm$^{-3}$.

The seed of the perturbation is obtained using a sinusoidal function for the $v_y$ component of the velocity field. For the initial velocity, we then have

\begin{equation}
 v_x(y)=v_2+(v_1-v_2)~f(y)\,,\qquad
 v_y(x)=\Delta v_y\sin{(n\pi x)}\,,
\end{equation}

\noindent
where we assume $n=2$ and $\Delta v_y=0.1$~cm.s$^{-1}$. Also note that the $v_x$ component has been smoothed using the same ramp function used for the density, and that $v_1=~0.5$~cm.s$^{-1}$ and $v_2=-0.5$~cm.s$^{-1}$, which corresponds to the high and low density bands respectively.

Figure \ref{figure4} shows four snapshots of the growth of the Kelvin-Helmholtz instability at different times for the calculation using {\sl IAD}$_0$ (first row) and the standard SPH implementation (second row). As can be seen, the standard formulation does a poor job of resolving the structure of the instability. Although in the beginning the main shape is in gross agreement with the tensor simulations, the final image is blurry, the interface is not well-defined and the shape is incorrect. In the case of the tensor calculation, the instability grows cleanly and at good rate, and the definition of the extremes of the billows, where the finest structure appears, is clearly enhanced. To achieve similar results to those obtained with the {\sl IAD$_0$}~ technique, different methods have been proposed to maintain the standard description, mainly based on including an artificial thermal conductivity \citep{price08}. This method provides reliable results, but includes a new set of parameters and estimates, such as maximum signal velocity between two particles to obtain a diffusion parameter. Furthermore, the inclusion of an artificial thermal conductivity leads to an extra dissipation of gradients away from discontinuities, hence it needs to design some means of controlling the amount of dissipation. 

In Fig. \ref{figure5}, we present the evolution of total linear momentum during the development of the instability. Although both schemes give a very satisfactory momentum conservation, we can see that conservation using the {\sl IAD$_0$} method is at least one order of magnitude better than that of standard scheme.  

To test the tensor approach in a more hostile scenario, we diminished the value of the amplitude of the initial velocity perturbation by an order of magnitude (i.e. $\Delta v_y=0.01$~cm.s$^{-1}$). In the third and fourth rows of Fig. \ref{figure4} we show the results of the simulations using the configurations {\sl IAD$_0$} and {\sl STD}. It is clear that in the standard formulation the instability \textit{was unable} to grow, while it does using \textit{IAD$_0$}. Changing the geometry of the initial particle distribution from a square to a uniform hexagonal lattice or reducing the size of the smoothing length taking $n_b=36$~neighbors did not appreciably change the results. We can then state, that the matrix approach has inherent virtues that are absent in the standard formulation, mainly related to the generalization of the derivation technique to a tensor expression, which, in some sense, diminishes the errors derived from the discretization.

It is also worth to mention that during the simulations carried out using the standard calculation a generalized clumping of particles in the low density regions often appeared. This is a well-known problem, called pairing or tensile-instability, where particles get ``stuck" if $r_{ij}<\frac{2}{3}~h$ owing to an unstable stress-strain relation that occurs typically when the second derivative of the kernel becomes negative and high strain is developed between particles. This problem eventually leads to difficulties in resolving the distances between neighbors, incorrect interpolations, and finally large non-physical behavior of the particles that halts the calculation. A typical way to cope with this problem is to adaptively change the kernel slope so that it becomes more centrally peaked \citep{cabezon08}. However, the {\sl IAD}$_0$ method was unaffected by this problem, hence we found that the tensor method seems to be less susceptible to pairing instability. The ability of matrix methods to cope with pairing instability has been reported by other authors \citep{oger07}, and they merit further investigation in the context of {\sl IAD}$_0$.

\subsection{Rayleigh-Taylor instability}

The simulation of the growth of the Rayleigh-Taylor instability in a stratified fluid in the presence of gravity is a classical test of subsonic fluid dynamics. In its simplest form, a box containing two fluids with different densities separated by a sharp transition region is placed inside a gravitational field \citep{youngs84}. If the denser fluid is located on top of the lighter, the system becomes physically unstable and any small perturbation of the interface leads to fluid overturn. The denser fluid sinks and the lighter one rises, producing characteristic structures known as spikes and bubbles respectively. After a linear phase of growth, which can be studied analytically, the system enters a complex non-linear regime. In real fluids, however, the viscosity of the fluid prevents the growth of the instability at short wavelengths, thus viscosity tends to dampen or even to completely suppress the instability. One serious problem of numerical schemes incorporating the artificial viscosity formulation is that they introduce too much viscosity into the system, compromising the growth of instabilities. As we later show, the integral approach to the derivative, Eqs. (\ref{momentumLqij}) and (\ref{energyqij}), provides a substantial improvement in the simulation of the RT phenomenon when the initial perturbation amplitude is small.

A sample of N=62\,500 particles was distributed in a squared lattice of 1 cm side. The mass of the particles was conveniently arranged to reproduce the density profile given by

\begin{equation}
\rho(x,y)=
\begin{cases}
\rho_2 & y \ge \Delta y \,\\
(\frac{\rho_2-\rho_1}{2\Delta y})~y + (\frac{\rho_1+\rho_2}{2})& -\Delta y <  y < \Delta y \,\\
\rho_1 & y \le -\Delta y\,
\end{cases}
\end{equation}

\noindent where $y$ is the vertical coordinate with the origin at the interface and $\Delta y$ is the width of the transition zone between both fluids. The density in the two zones were set to $\rho_1=~1$~g~cm$^{-3}$ and $\rho_2=2$~g~cm$^{-3}$ respectively, and the size of the transition zone was taken to be $\Delta y=0.05$~cm.

The integration of the hydrostatic equilibrium equation gives the pressure distribution along the fluid

\begin{equation}
P(x,y)=
\begin{cases}
-g~\rho_2~(y-\frac{1}{2})\,;& y\ge \Delta y \,\\
\begin{split}
-g&\left[\rho_2~(y-\frac{1}{2})+\right.\\
&(\frac{\rho_2-\rho_1}{4\Delta y})(y^2-{\Delta y}^2)+\\
&\left.(\frac{\rho_1+\rho_2}{2}) (y-\Delta y)\right]\,;\end{split}& -\Delta y< y < \Delta y\,\\
g(\frac{\rho_2}{2}-\rho_1 y)\,;& y\le -\Delta y\,
\end{cases}
\end{equation}

\noindent where g is the gravity, here of value g=0.5~cm~s$^{-2}$. With this setting, the speed of sound at the inter-phase is $c_s=0.7$~cm~s$^{-1}$. A perturbation in the boundary layer with $\lambda=0.25$~cm was seeded by giving an initial small vertical velocity $v_y$ to the particles according to the prescription \citep{abel11}

\begin{equation}
v_y(x,y)=\frac{\Delta v_y}{4}\left\{1+\cos\left[8\pi\left(x+\frac{1}{4}\right)\right]\right\}\left\{1+\cos\left[5\pi\left(y-\frac{1}{2}\right)\right]\right\}\,,
\end{equation}

\noindent and $v_y$ was set to zero for $y$ positions above 0.7 cm and below 0.3 cm. The value of velocity perturbation was set to $\Delta v_y=0.01$~cm~s$^{-1}$. In Fig. \ref{figure6} we show several snapshots of the development of the Rayleigh Taylor instability calculated using {\sl IAD$_0$}. As we can see, the RT instability is able to grow after a few tenths of a second. When the same calculation was attempted using the standard SPH scheme to compute the acceleration, the results were of much lower quality, the instability was totally damped and there was no development of the RT fingers. As in the case of the Kelvin-Helmholtz instability, changing the initial particle setting from the square to an hexagonal regular lattice or reducing the size of the smoothing length did not qualitatively alter the above conclusions. A similar conclusion was reported by \citet{abel11} who used a variant of standard SPH, albeit not fully conservative, to simulate the grow of the RT instability using a similar initial setting of the experiment. This negative result does not mean that standard SPH is unable to simulate this phenomena because it could handle it after a careful initial setting that minimizes the numerical noise. Nevertheless, the results of our simulations indicate that the tensor method is less dependent on the particular geometry of the initial lattice.
  
In Fig. \ref{figure7} we show the evolution of the center of mass of the spikes $y_s$ from the initial contact surface. According to the theory, the evolution is exponential during the initial stage $\xi(t)=~\xi_0\exp\left[\Gamma (t-t_0)\right]$, roughly until the initial perturbation $\xi_0$ at $t_0$ (here we took $t_0=1.1$~s) has grown to a size $\xi_l\simeq 1/k$ where $k$ is the wave number. In our tests, $\xi_l\simeq 0.04$~cm. The theoretical growth rate $\Gamma_t$ during the exponential phase is given by $\Gamma_t=\sqrt{A_t~k g}$, where $A_t=(\rho_2-\rho_1)/(\rho_2+\rho_1)$ is the Atwood number. The exponential phase is followed by a brief second stage in which the penetration of the dense fluid evolves as $y_s\propto g t^2$~until drag  takes over and both falling spikes and rising bubbles reach a limiting speed. According to Fig. \ref{figure7}, all these features are present in the simulations. Nevertheless, there was only a qualitative agreement with the analytical growth rate during the initial exponential phase because the value of $\Gamma$~inferred from the simulations was almost one half of the theoretical value. This is unsurprising because a reliable quantitative estimation during the initial stage needs both a higher resolution as well as a better initial setting than that we are currently using. In particular, the transition zone between light and dense fluids has to be sharper to correctly represent the Atwood number in this stage. An analytical expression for the terminal velocity of a rising bubble in a tube was obtained by \citet{Layzer55}, $v_b=0.51 \sqrt{A_t g l}$~where $l$ is the radius of the tube. Taking $l=\lambda=0.25$~cm, we get  $v_b=0.104$~cm.s$^{-1}$~in good agreement to the numerical value deduced from Fig. \ref{figure7}.

\subsection{The wall heating shock test}

We now consider simulations of supersonic events. In the test known as the wall shock problem, a spherical or cylindric supersonic stream of gas is launched towards its geometrical center forming a highly compressed region. For these geometries, simulations can be compared to the predictions of an analytical approach to the evolution of thermodynamic variables as a function of the initial conditions \citep{noh87}. Although the gross features of the event are correctly captured by SPH simulations, it is well-known that schemes based on artificial viscosity have difficulties in providing a detailed description of the wall heating shock test. The reason is that AV spreads the shock over several computational cells, which induces a non-physical increase in the internal energy ahead of the shock. For converging flows, a large artificial spike in internal energy is observed at the convergence center. As a consequence, a profound dip in the density profile appears to keep the pressure smooth. However, the inclusion of a good amount of artificial viscosity is mandatory to providing a successful description of the shock.

Our purpose behind this test was to discern whether the integral route to calculating derivatives can provide more reliable results than the standard procedure. Therefore, we use {\sl IAD$_0$} and Eqs. (\ref{density}), (\ref{momentumLqij}) and (\ref{energyqij}) to carry out the simulations.

A sample of N=57\,600 particles of the same mass were evenly spread across a lattice to ensure that the density was homogeneous, $\rho=1$~g~cm$^{-3}$. The initial pressure and internal energy of particles were negligible. The system was imploded by imposing a spherically symmetric velocity field $v_r=-1$~cm~s$^{-1}$. In Fig. \ref{figure8} we show the density and radial velocity profiles at t=0.3~s, when the shock is well-developed. Both profiles follow a simple pattern that remains close to the analytical estimations. In the central region, $r<0.1$~cm, a plateau of compressed material forms, with a characteristic density of $\rho_s\simeq 14.5$~g~cm$^{-3}$ a little lower than the analytical value of $\rho=16$~g~cm$^{-3}$ for $\gamma=5/3$ \citep{noh87}. In this inner region matter remains stagnated with a velocity close to zero. Outside the plateau the density abruptly declines through the shock front trying to regain its initial value, whereas the velocity increases to reach $v_r=-1$~cm~s$^{-1}$ at the incoming, but still unshocked, matter. As we can see in Fig. \ref{figure8}, the simulation using the tensor approach is of higher quality than that of standard derivatives. In particular, the numerical oscillations in the post-shock region in both, density and velocity, were considerably smaller in the tensor formulation. Nevertheless, the dip in the density profile and the maximum value achieved by density were similar in both calculations.

\subsection{Sedov test}

In the Sedov test, the evolution of a shock wave front born as a consequence of a point-like explosion is studied as it propagates in a homogeneous medium. The problem of an intense explosion in a gas is a standard test for hydrocodes that is of relevance to astrophysics, where it is not rare to find strong shocks in many scenarios involving fluid motions at high velocity. The theoretical solution was found by L.I. Sedov by applying self-similar methods and dimensional analysis for different geometries \citep{sedov59}. In its simplest formulation, the Sedov problem has an initially cold gas at rest. At t = 0 s there is a point explosion at the origin that in \citet{sedov59} was treated as an instantaneous release of energy at the origin, and assumed that the background material through which the expanding gas sweeps behaves like a perfect gas. For these initial conditions there are precise, albeit algebraically complicated, analytical expressions for the fluid variables. In the case of SPH, the artificial viscosity smears the shock over 2-3 times the smoothing-length. As a consequence, the density jump across the shock front is always smaller than the factor of four predicted by the theory for $\gamma=5/3$. In more than one dimension, resolution issues are crucial not only to resolve the peak of the blast wave but also to reproduce the correct post-shock variables downstream and the structure of the rarefied tail close to the origin. We want to investigate the extent to which the use of the integral approach to the derivative can improve the results of the SPH simulations.

The initial setting was similar to that of the wall heating test but this time the velocity was set to zero everywhere at t=0 s, to ensure that the initial configuration is in equilibrium. To create the explosion, a large amount of internal energy was instantaneously released in a small region around the center of the box. To smooth the initial discontinuity, we took an initial pressure step that decays as a Gaussian function

\begin{equation}
P(r)=P_2+(P_1-P_2)\exp\left[\frac{-r^2}{\sigma^2}\right]\,,
\end{equation}

\noindent where $P_1$ and $P_2$ are the pressures in zones left and right of the step and $\sigma$ sets the width of the pressure decay. In the present simulations, $P_1=10^4$~dyn~cm$^{-2}$ and $P_2=1$~dyn~cm$^{-2}$ and $\sigma^2=36$~cm$^{2}$, which smooths the internal energy step over about 5-6 times the smoothing length. In Fig. \ref{figure9} we depict the density and velocity profiles once the self-similar state has been achieved. As we can see  the profiles calculated using the tensor and standard approaches to calculating gradients do not differ so much. The peak in the density profile is about 80\% of what is expected from a strong shock moving through a $\gamma=5/3$~gas but the profile in the post-shock zone is smoother in the calculation using {\sl IAD$_0$}. The velocity profile is also similar but the tensor calculation showed fewer oscillations and a more ordered behavior in the post-shock tail. These results agree with our main conclusions given in the previous test dealing with the wall heating shock, reinforcing the idea that for identical initial settings {\sl IAD$_0$} provides more reliable results than standard derivatives also for supersonic events. 

\section{Handling linear phenomena: momentum and energy equations using IAD}
\label{section5}

Ensuring both exact linear interpolation and perfect momentum and energy conservation using either full {\sl IAD} or other similar schemes, seems to be difficult. In this respect, the best physical systems to try to conciliate both things  are pure linear systems for which linear acoustic wave propagation is a representative example. A suitable alternative SPH formulation can be found starting from the differential acceleration equation  

\begin{equation}
{\bf \ddot {r}}=-\frac{1}{\rho}{{\bf \nabla}}P=-\left[{\bf \nabla}\left(\frac{P}{\rho}\right)+\frac{P}{\rho^2}{\bf \nabla}\rho\right]\,,
\label{SPHmomentum0}
\end{equation}

\noindent which, using Eq. (\ref{nablafSPH}), can be easily accommodated within the SPH formalism

\begin{equation}
{\bf \ddot {r}}_a=-
\sum_b m_b\left(\frac{P_a}{\rho_a^2}+\frac{P_b}{\rho_b^2}\right){\bf \nabla}\tilde W_{ab}\,.
\label{SPHmomentum}
\end{equation}

This equation can be deduced independently using the variational principle and, provided the kernel is symmetrized $\tilde W_{ab}=0.5(W_{ab}(h_a)+W_{ab}(h_b))$, it conserves linear and angular momentum exactly. If instead of Eq. (\ref{nablafSPH}) we use Eq. (\ref{nablaf}) to compute Eq. (\ref {SPHmomentum0}), the resulting equation is

\begin{equation}
\begin{split}
{\bf \ddot{r}}_a =&-\left[{\bf \nabla}\left(\frac{P}{\rho}\right)+\frac{P}{\rho^2}{\bf \nabla}\rho\right]=\\
&-\sum_b m_b\left(\frac{P_a}{\rho_a^2}+\frac{P_b}{\rho_b^2}\right){\bf \nabla}\tilde W_{ab}+2~P_a \sum_b \frac{m_b}{\rho_a~\rho_b}{\bf \nabla}\tilde W_{ab}\,.
\end{split}
\label{momv3}
\end{equation}

This particular form of the acceleration equation has several interesting features:  1) In the continuum limit it reduces to the standard expression, Eq. (\ref{SPHmomentum}), because the last term in the equation vanishes. 2) As shown in Sect. 3.2, an isobaric system with smooth density gradients remains in equilibrium with negligible acceleration for a long time (see Fig. \ref{figure3}).
 
However, it is also evident that the inclusion of the last term in Eq. (\ref{momv3}) breaks its symmetry. Still, in the linear limit, momentum is conserved to a high extent as can be easily demonstrated in 1D. We first expand the magnitudes $P/\rho^2$ in Eq. (\ref{momv3}) assuming a smooth density gradient

 \begin{equation}
\frac{P_a}{\rho_a^2}\simeq \frac{P_a}{\rho_a\rho_b}\left(1-\frac{\Delta\rho}{\rho_b}\right)\,, \qquad
\frac{P_b}{\rho_b^2}\simeq \frac{P_b}{\rho_a\rho_b}\left(1+\frac{\Delta\rho}{\rho_a}\right)\,,
\label{expansionro}
\end{equation}

\noindent where $\Delta\rho=\rho_a-\rho_b$ and $\vert\Delta\rho\vert/\rho << 1$. Inserting Eq. (\ref{expansionro}) into Eq. (\ref{momv3}) and neglecting higher order terms, the acceleration is given by

\begin{equation}
\ddot x_a=-\frac{\sum_b\frac{m_b}{\rho_b}\left(\frac{P_b-P_a}{\rho_a}\right)(x_b-x_a) W_{ab}}{\sum_b\frac{m_b}{\rho_b}(x_b-x_a)^2 W_{ab}}\,.
\label{lineariad}
\end{equation}

\noindent If we assume a linear pressure profile within the kernel domain of the particle

\begin{equation}
{P_b}= P_{a}+\left(\frac{\partial P}{\partial x}\right)_a (x_b-x_a)\,,
\label{expansion}
\end{equation}

\noindent the acceleration becomes

\begin{equation}
\ddot x_a=-\frac{1}{\rho_a}\left(\frac{\partial P}{\partial x}\right)_a \,,
\label{momentumX}
\end{equation}

\noindent which is the one-dimensional Newton equation, thus leading to an exact value for the acceleration. In consequence, global momentum is well-preserved provided that the system remains in the linear regime and the density gradients are smooth. According to this discussion, we complete the acceleration equation obtained in Sect. \ref{section3} using the {\sl IAD$_0$} scheme with a corrective term    

\begin{equation}
\begin{split}
\ddot x_{i,a}=-\sum_{b=1}^{nb} m_b&\left(\frac{P_a}{\Omega_a\rho_a^2}\mathcal A_{i,ab}(h_a)+\frac{P_b}{\Omega_b\rho_b^2}\mathcal A'_{i,ab}(h_b)+\right.\\
&\left. \Pi_{ab}~\widetilde{\mathcal A}_{i,ab}-\frac{\Psi P_a}{\rho_a\rho_b}\mathcal A_{i,ab}(h_a)\right)\,,
\end{split}
\label{momentumLcomplete}
\end{equation}

\noindent where $0\le\Psi\le 2$ is a parameter controlling the strength of the applied correction. For $\Psi=0$, the already checked {\sl IAD$_0$} scheme results, whereas for $\Psi=2$ we have full linear {\sl IAD}.
 
To take into account the corrective term in the energy equation we include the power released/absorbed by such term in Eq. (\ref{energyqij}) 

\begin{equation}
\begin{split}
\left(\frac{du}{dt}\right)_a=\sum_{b=1}^{n_b}\sum_{i=1}^d &m_b(v_{i,a}-v_{i,b})\cdot\left(\frac{P_a}{\Omega_a\rho_a^2}~\mathcal{A}_{i,ab}(h_a)+\right.\\
&\left. \frac{\Pi_{ab}}{2}~\widetilde{\mathcal A}_{i,ab}(h_b)-\frac{\Psi P_a}{\rho_a\rho_b} v_{i,a}\mathcal A_{i,ab}(h_a)\right)\,.
\end{split}
\label{energycomplete}
\end{equation}

\noindent such that total energy is exactly conserved. For very subsonic movements, the energetic contribution of the correction term is small. For $\Psi=0$, Eqs. (\ref{momentumLcomplete}) and (\ref{energycomplete}) reduce to the fully conservative {\sl IAD$_0$} scheme. Alternatively, the case of $\Psi=2$ could be used to handle linear systems with smooth density gradients as it shows enhanced interpolation abilities and good conservative properties. 

Sound wave propagation is an example of a system that evolves in the linear limit of Euler equations. It is therefore a test ideally suited to highlighting the potential advantages of the {\sl IAD} scheme. A homogeneous system with $\rho_0=~1$~g~cm$^{-3}$ was simulated using N=62\,500 particles uniformly distributed in a lattice of size 1 cm. The initial values of both pressure and density were set to one. A sample of n$_p=200$ particles inside a circle of radius $R_p=0.032$~cm located at the center of mass of the lattice was obliged to oscillate with small amplitude and period P. This compact set of particles emulated the effect of an external oscillating {\sl piston}~moving adiabatically onto the initially unperturbed gas. As the {\sl piston} moves, it launches regular trains of circular waves that propagate at the sound speed $c_s=\sqrt{\gamma P/\rho}=1.29$~cm~s$^{-1}$~for $\gamma=5/3$. We want to check the ability of the different aforementioned schemes to handle with this problem.

Particles belonging to the {\sl piston} move homologously, following the harmonic oscillator law

\begin{equation}
 v_r(r,t)= v_r^{m}(r) \cos(\omega t)\,,
\label{harmonic}
\end{equation}

\noindent where $\omega$ stands for the angular frequency, and the maximum value of the radial velocity of a particle located at distance $r$ from the center is

\begin{equation}
 v_r^m(r)= v_r^m (R_p)\left(\frac{r}{R_p}\right)\,.
\label{radialvel}
\end{equation}

The resolution places limits on both the period and the maximum velocity at the {\sl piston} head $v_r^m (R_p)$. The value of the period was set to P=0.05 s so that around eight complete waves were launched before the first one reached the limits of the system. The value of the maximum velocity at the piston head was set to $v_r^m (R_p)=0.16$~cm~s$^{-1}$. 

The results of the simulations for the three SPH approaches: {\sl IAD}, {\sl IAD$_0$} and Standard are summarized in Fig. \ref{figure10}, which represents the radial velocity profile of the gas  at time t=0.35 s. It is quite evident that the spherical symmetry was better preserved during the {\sl IAD} calculation, while there is a progressive degradation of the symmetry in the {\sl IAD$_0$} and standard calculations respectively. The worst case corresponds to the standard scheme but even in this case spherical symmetry is reasonably preserved. A more quantitative analysis can be done making use of the analytical solution for diverging circular waves, usually known as waves in a membrane in the literature. The solution for free harmonic traveling waves in a circular membrane is written as

\begin{equation}
A(r,t)= A_{m} J(u) cos(wt)\,,
\label{membrane}
\end{equation} 

\noindent where $u=kr$, $k$ is the wave number and $J(u)$ is the Bessel function with radial and angular modes m=n=0. The Bessel functions are constrained by $J(0)=1$ at the center of the wave train. The profile for the radial velocity given by Eq. (\ref{membrane}) with $A_m=v_{r}^m(R_p)$ and the comparison with the numerical profiles obtained using the different schemes are also shown in Fig. \ref{figure10}. The analytical solution reminds one of a damped cosine wave propagating at the sound speed  $c_s=1.29$~cm.$^{-1}$. On the whole, the three schemes were able to correctly track the evolution of the waves. However, a close inspection reveals small differences among them. The best value of wave velocity corresponds to the {\sl IAD$_0$} scheme ($c_s^{iad_0}=1.285$~cm.s$^{-1}$) followed by the standard ($c_s^{std}=1.267$~cm.s$^{-1}$), whereas full {\sl IAD} falls a bit short ($c_s^{iad}=1.25$~cm.s$^{-1}$). Even though the damping in the three schemes is always larger than that of the analytical solution, it seems that the worst case also corresponds to the fully {\sl IAD} scheme built setting $\Psi=2$ in Eqs. (\ref{momentumLcomplete}) and (\ref{energycomplete}).
 
The simulation of linearized systems represents the most favored situation to apply exact linear interpolation schemes such as {\sl IAD} because, as shown above, conservation of momentum and energy is probably as good as in the standard formulation. Nonetheless, the numerical experiments with the acoustic waves suggested limited advantages of using these schemes. In fact, 
some details of the wave evolution are better represented by {\sl IAD$_0$} or even by the standard scheme. Only the spherical symmetry was more accurately preserved in the experiments using {\sl IAD}.  

\section{Applying {\sl IAD$_0$} to astrophysics: the stability of a Sun-like polytrope}
\label{section6}

Applying Eqs. (\ref{density}), (\ref{momentumLqij}) and (\ref{energyqij}) to astrophysics is straightforward by just adding the gravitational acceleration to the momentum equation. We approached gravity using a multipolar expansion of the force \citep{hern89}, up to quadrupole contributions. The 3D structure of a Sun-like polytrope was simulated using a sample of $N=10^5$~particles with equal mass. To achieve the equilibrium, we proceeded in three stages. First, the radial coordinate of each particle was set following the 1D density profile of the polytrope with its angular position determined at random. In the second stage, we allowed the particle sample to relax under the action of pressure and gravity forces but the movement of the particles was constrained to keep their distance to the center constant. In a third stage, we allowed the particle sample free to approach the equilibrium configuration. We followed the third stage using both {\sl IAD$_0$} and STD schemes with the AV coefficients set to $\alpha=1$, $\beta=2$. The main goal of this test is to show that {\sl IAD$_0$} (and probably other matrix methods) can be used to simulate astrophysical scenarios with good results, excellent conservation properties and with low computational penalty. 

In Fig. \ref{figure11} (upper row) we depict the evolution of density for two particles initially located at the center and the surface of the star respectively. As we can see, these particles display the typical oscillatory behavior in both regions. As expected, the amplitude of the oscillations is small close to the center and larger at the surface. The relaxation towards equilibrium is slower for the tensor method but the structure of the star after $t=2$~h of relaxation is very similar. With the exception of a tiny region close to the surface, we verified that the gradient of pressure matched gravity along the polytrope in both calculations. In Fig. \ref{figure11} (lower row) we show the evolution of the kinetic energy and the instantaneous position of the center of mass during the relaxation process. The evolution is similar in both calculations but the level of kinetic energy is always a bit higher when {\sl IAD$_0$} is used. The energy conservation after two hours of evolution is rather good for both schemes $\simeq 3~10^{-5}$. In the bottom right of Fig. \ref{figure11} we show the evolution of the moduli of the center of mass position. We see that momentum is not perfectly conserved during the simulation in both simulations, mainly owing to the multipolar approach to the gravitational force. Nevertheless, the tensor scheme provides a better conservation of momentum than the standard SPH calculation.

The higher level of kinetic energy and the slower relaxation rate towards complete mechanical equilibrium shown by {\sl IAD$_0$} suggest that for disordered systems the numerical noise could be higher for the tensor method. As shown in Fig. \ref{figure11}, the recalculation of the evolution of the polytrope using larger values of the AV coefficients, $\alpha=1.5,\beta=3$~led to a significant reduction in the level of spurious kinetic energy, while the evolution of the density at the center and surface of the star remained unaltered. This implies that the numerical noise is the main agent responsible for the excess of kinetic energy seen during the relaxation process.
          
To explore the performance of the algorithm, the tolerance parameter $\theta$, controlling the accuracy of the multipolar expansion, was varied from $0\le \theta\le 1$, where $\theta=0$ corresponds to direct particle to particle interaction. This benchmarking analysis is summarized in Fig. \ref{figure12}. We see that for standard values of the tolerance parameter, $\theta\simeq 0.7$, widely used in current SPH calculations, the computational overload is around $40\%$. This overload rapidly decays as $\theta\rightarrow 0$, as expected. However, the number of operations in the multipolar calculation of gravity scales as $\simeq N log N$ \citep{hern89}. We therefore expect a reduction in the computational overload with increasing numbers of particles. Nevertheless, we note that all these calculations were performed using a serial code. An invaluable property of SPH is that only a single tree-walk is needed to find the neighbors (and calculate gravity if needed). Once this information is available the remaining calculations (density, momentum and energy equations, {\sl IAD$_0$} terms) can be performed with linked-lists that can be directly parallelized. Following that idea, we found that for a 3D simulation of $2\cdot 10^5$ particles in a standard 4-core desktop computer, the gravity calculation (which remains serial) takes around 30 times more wall-clock time than the rest of the calculations together (parallelized with OMP using 4 cores). Knowing this, the computational overload due to the {\sl IAD$_0$} calculation remains sub-dominant.

\section{Conclusions}

We have presented and verified a novel procedure ({\sl IAD}) to evaluate gradients in the context of SPH simulations. The mechanism for calculating derivatives relies on an integral approach, which ensures that the derivative of a linear function is exactly obtained, even after transforming integrals to summations. The drawback is the greater algebraic complexity because gradients are estimated using the tensor expression given by Eq. (\ref{matrix}). Nevertheless, our new scheme is not significantly more complex than the standard one, easy to implement and includes the classical formulation as a particular case. A shortcoming of our new method is that it leads to non-conservative movement equations. Thus, we have developed and tested a restricted interpretation of the method, referred to as {\sl IAD$_0$}, which sacrifices exact linear interpolation to ensure a perfect conservation of momentum and energy. The analytical considerations and numerical experiments described in Sects. \ref{section2} and \ref{section4} strongly indicate that {\sl IAD$_0$} behaves better than the standard method in computing gradients. The modified momentum equation obeying these derivative rules was developed in Sect. \ref{section3}, resulting in Eq. (\ref{momentumL}). Since the movement equation was obtained from the Euler-Lagrange variational principle assuming isentropic evolution, the ensuing equation was fully conservative and explicitly included the $\nabla h$ terms. A conservative energy equation compatible with the momentum equation was also developed. To handle with shocks the scheme was completed by including of the standard artificial viscosity formalism resulting in Eqs. ({\ref{momentumLqij}) and (\ref{energyqij}), which in addition to the density equation in Eq. (\ref{density}), summarizes the mathematical formalism linked to {\sl IAD$_0$}. 

The formulation of SPH using matrix methods based on the variational approach \citep{bonet99} has been used in CFD to successfully simulate a variety of problems (generally in two dimensions) from fluids to the impact and fracture of solid bodies. Nevertheless, none of these schemes are able to simultaneously achieve re-normalization, exact momentum and energy conservation, perfect linear interpolation and implicitly include the gradient of the smoothing length in the equations \citep{oger07}. Nevertheless, {\sl IAD$_0$} is probably the optimal formulation because it fulfills almost all the above requirements with a moderate computational overload.

Four tests were performed in two dimensions to verify the performance of the method (see Table~1), two of them in connection to bi-dimensional subsonic hydrodynamic evolution (KH and RT instabilities) and the other two related to the description of highly supersonic phenomena such as the wall heating shock and the Sedov tests. In all cases the performance of the new scheme was superior, although in the supersonic numerical experiments the improvement was modest. The tests of the growth of the Kelvin-Helmholtz and Rayleigh-Taylor instabilities clearly showed the power of the {\sl IAD$_0$} scheme because no growth at all was seen when the standard scheme was used with small initial perturbations, whereas instabilities were able to grow when {\sl IAD$_0$} was used. This does not of course mean that the standard SPH technique cannot give a satisfactory answer to these problems but simply states that for the same initial conditions the tensor method seems to be more stable and sensitive to small perturbations. Therefore, these results indicate that simulations using the {\sl IAD$_0$} scheme are less dependent on the initial setting of the particles. As expected of a tensor method, this advantage could probably be reinforced as the dimensionality increases.

In Sect. \ref{section5} we devised and discussed a slightly different formulation, which combines exact linear interpolation and good conservative properties, given by Eqs. (\ref{momentumLcomplete}) and (\ref{energycomplete}). The main point here is that the fully {\sl IAD}  method would be only useful for describing linearized systems with smooth density gradients. Nevertheless, a single parameter allows us to easily switch from {\sl IAD} to {\sl IAD$_0$} and in this sense the formulation given in Sect. \ref{section5} is more general. As sound waves are the prototype of a linearized system we have used the new scheme to simulate the 2D propagation of acoustic waves. The simulations indicate that there are no particular advantages of using {\sl IAD}~instead of either {\sl IAD$_0$} or standard SPH, because only the symmetry was better preserved during the evolution of the wave trains.

A preliminary application of {\sl IAD$_0$} to astrophysics was discussed in Sect. \ref{section6} in connection to the stability of a Sun-like star described by a polytropic EOS. The main goal was to demonstrate that matrix SPH methods can be used to handle 3D self-gravitating bodies with satisfactory results and to explore the real computational overload when long-range forces need to be computed. For a reasonable accuracy in the calculation of gravity, we have estimated a computational overload $< 50\%$ for a serial code, with respect to that of standard SPH, when a multipolar expansion is used to calculate the gravitational force. The overload becomes negligible for a precise particle-to-particle calculation of gravity scaling as $N^2/2$.

To summarize we could say that simulations carried out using the SPH scheme obtained with the {\sl IAD$_0$} approach to the gradients always led to improved results with respect to standard SPH. As the new scheme is both fully conservative and more precise in making interpolations, it could be an alternative to the standard technique in handling systems subjected to small perturbations. This conclusion is supported by the results of our numerical study of the growth of Kelvin-Helmholtz and Rayleigh-Taylor instabilities. In addition, our simulations of supersonic phenomena also improved when the tensor approach was used. The main drawback of this method is that it increases the computational overload, but one has to keep in mind that there is not always a linear relationship between algebraic complexity and computational charge. This could be true for hydrocodes that incorporate time-consuming physics or when the SPH algorithm built with {\sl IAD$_0$} allows longer time steps. Even more, using linked-lists a direct parallelization of the method is possible as the calculations are carried out of the tree-walk, keeping the computational overload of the {\sl IAD$_0$} almost negligible compared to more time-consuming sections of the code. 

Although the presented results are encouraging more work needs to be done to confirm and extend the conclusions of our proposal. For the most part, the tests presented in this work to validate the scheme were carried out in 2D boxes using well-ordered initial models. The simulation of realistic astrophysical scenarios generally involves, however, a quite different numerical setting. Many of these calculations have to be performed in 3D using a random-like initial particle distribution and incorporating the long-range gravitational force. A detailed comparative analysis of the ability of {\sl IAD$_0$}~and standard SPH to cope with these scenarios is beyond the scope of the present work and is left to a forthcoming publication. Nonetheless, we can suggest a couple of areas of difficulty in the tensor formulation that will probably come up in astrophysical 3D applications of the method: 1) As suggested in Sect. \ref{section6}, the tensor method might be more sensitive to the disorder of the particles and display a higher level of numerical noise than the standard scheme. 2) Free boundary conditions could be more difficult to handle using {\sl IAD$_0$} because the simplification hypothesis assumed in Eq. (\ref{approxI}) does not hold at the edge of the system. The difficulty in simulating sharp boundaries with matrix methods is a well-known problem of {\sl MLS}~schemes. According to \citet{oger07}, it can be solved by taking special conditions at the limits of the system. This is not, however, a big concern in astrophysics because astronomical bodies never have abrupt boundaries. Thus, the stability test of the Sun-like star discussed in Sect. \ref{section6} does not reveal that the particles located close to the surface have peculiar behaviors. Although the initial imbalance between gravity and pressure gradient is more pronounced than in the standard formalism, the ensuing evolution towards equilibrium is not much different. 

\section*{Acknowledgements}

The authors acknowledge fruitful discussions with Antonio Rela\~no and Fayyaz Ahmad regarding the SPH technique. This work has been funded by the Spanish MEC grants AYA2010-15685, AYA2008-04211-C02-C01 and the Swiss Platform for High-Performance and High-Productivity Computing within the {\em supernova} project. It was also supported by DURSI of the Generalitat de Catalunya. The rendered SPH plots were made using the freely available $SPLASH$ code \citep{price07}.


\clearpage

\begin{figure}
\includegraphics[width=\columnwidth]{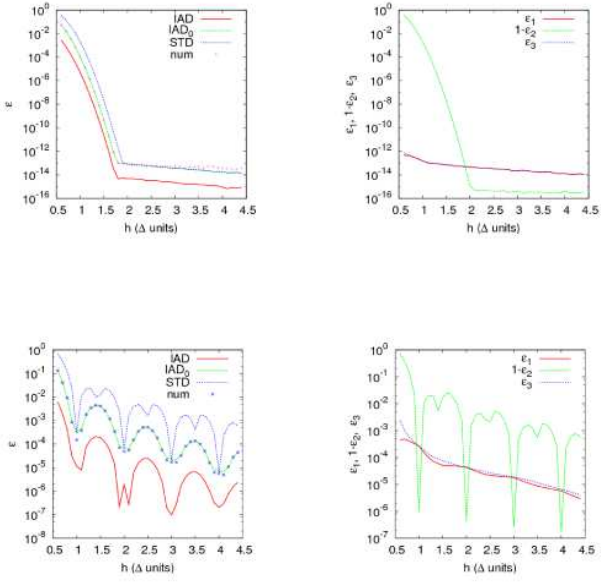}
\caption{Relative error $\epsilon$ in the first derivative of density $\rho(x)=(1+x)$~g~cm$^{-3}$ as a function of the smoothing length (in interparticle units) calculated using both tensor and standard SPH schemes. Crosses are for direct $\frac{\rho_b-\rho_a}{x_b-x_a}$ derivative estimation. Upper left is for the error when the Gaussian kernel is used. The upper right picture shows the contribution of  several error sources to the total error $\epsilon$. Bottom pictures are the same but for the cubic-spline kernel.}
\label{figure1}
\end{figure}

\begin{figure}
\includegraphics[width=\columnwidth]{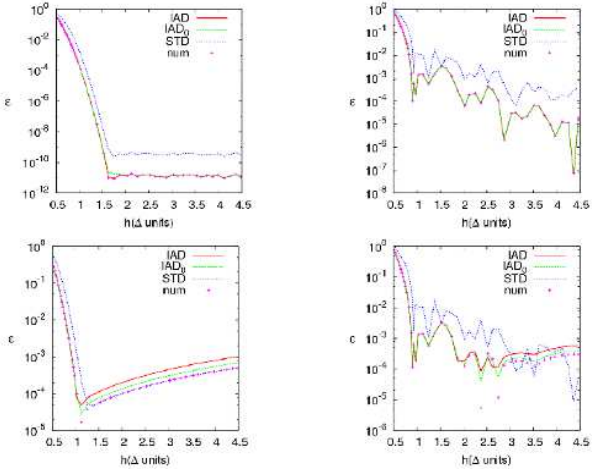}
\caption{Relative error $\epsilon$ in the first derivative of linear functions calculated in 2D as a function of the smoothing length. Upper rows depict the value of~$\epsilon$ for $\rho(x,y)=(1+x)$~g~cm$^{-3}$ obtained using the Gaussian (left) and the cubic spline (right) kernels. Bottom rows are the same but for the function $\rho(x,y)=1+\sqrt{x^2+y^2}$~g~cm$^{-3}$. Error profiles with crosses are for the direct $\frac{\rho_b-\rho_a}{r_b-r_a}$ derivative estimation.}
\label{figure2}
\end{figure}

\begin{figure}
\includegraphics[width=\columnwidth]{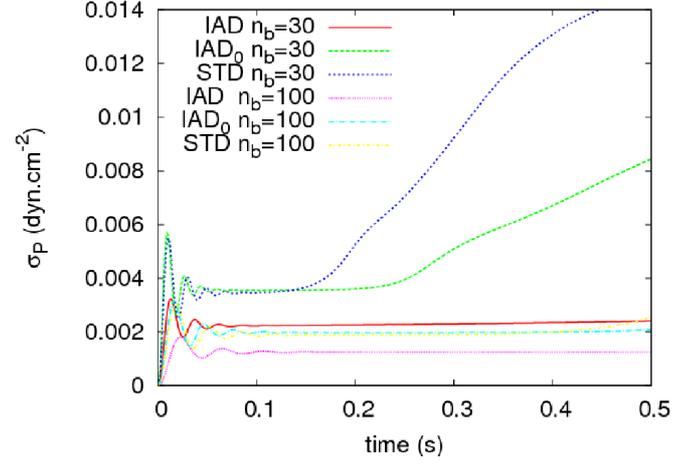}
\caption{Evolution of magnitude $\sigma_{P(t)}=\sqrt{\frac{\sum_b (P_b-P_0)^2}{N}}$ of an inhomogeneous 2D system, initially in hydrostatic equilibrium, calculated using the different SPH schemes mentioned in the text.}
\label{figure3}
\end{figure}

\begin{figure}
\includegraphics[width=\columnwidth]{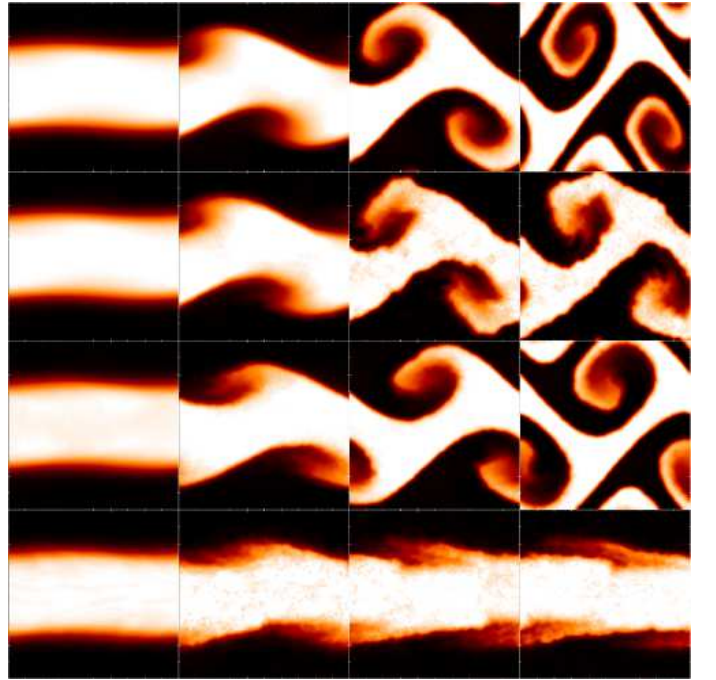}
\caption{Evolution of the Kelvin-Helmholtz instability. The snapshots show the color map of density at times $t=0.1,~1.0,~2.0$ and $3.0$~s for methods {\sl IAD$_0$} (first row) and standard SPH (second row), with $\Delta v_y=0.1$~cm~s$^{-1}$. Times $t=1.0,~3.0,~4.0$ and $5.0$~s for methods {\sl IAD$_0$} (third row) and standard SPH (fourth row), with $\Delta v_y=0.01$~cm~s$^{-1}$.}
\label{figure4}
\end{figure}

\begin{figure}
\includegraphics[width=\columnwidth]{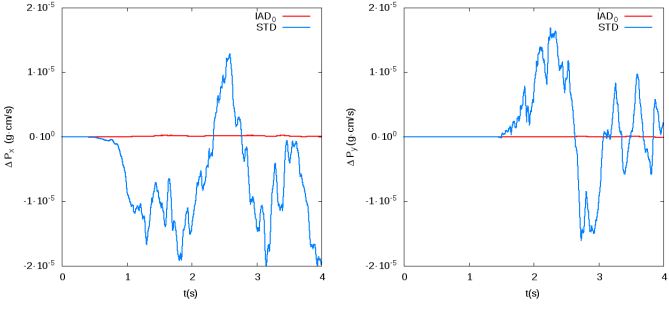}
\caption{Evolution of the absolute deviation from initial linear momentum during the development of the Kelvin-Helmholtz instability. Figure on the left is for the $x$~component of total momentum calculated using {\sl IAD$_0$} and standard (STD) schemes, whereas figure on the right is the same for the $y$~component.}
\label{figure5}
\end{figure}

\begin{figure}
\includegraphics[width=\columnwidth]{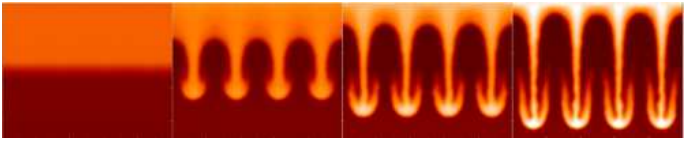}
\caption{Development of the Rayleigh-Taylor instability for Atwood number 1/3 calculated using the {\sl IAD$_0$} scheme. The snapshots show the color map of density at times $t=0.4,~4.2,~5.1$ and $5.7$~s.}
\label{figure6}
\end{figure}

\begin{figure}
\includegraphics[width=\columnwidth]{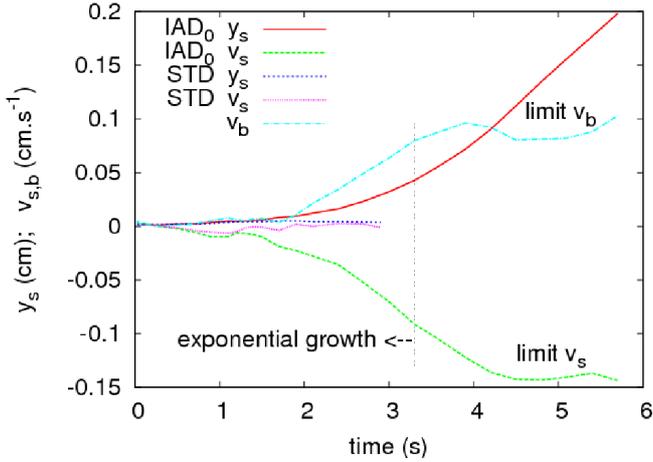}
\caption{Evolution of the center of mass of the spikes during the growth of the Rayleigh-Taylor instability depicted in Fig. \ref{figure6}. The velocity of the center of mass of the bubbles, $v_b$, and spikes, $v_s$, is also shown. For $t>4$~s, a limiting speed is reached.}
\label{figure7}
\end{figure}

\begin{figure}
\includegraphics[width=\columnwidth]{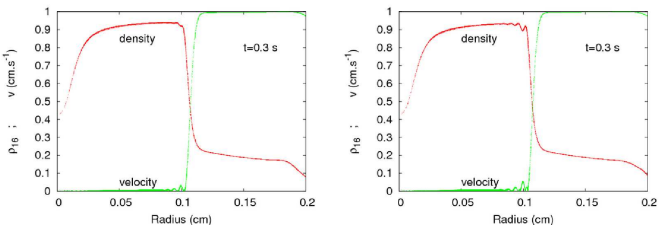}
\caption{Density and velocity profiles during the wall heating shock test. Figure on the left is for {\sl IAD$_0$} calculation, whereas figure on the right is for the simulation using the standard SPH scheme. Density is normalized to 16~g~cm$^{-3}$.}
\label{figure8}
\end{figure}

\begin{figure}
\includegraphics[width=\columnwidth]{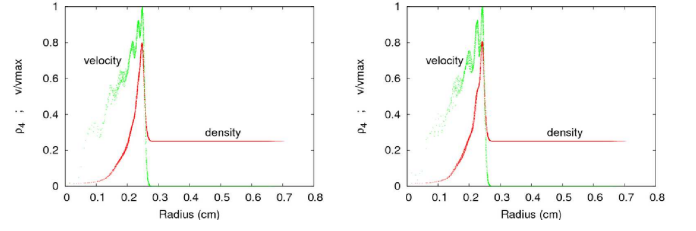}
\caption{Density and velocity profiles during the Sedov test. Figure on the left is for {\sl IAD$_0$} calculation, whereas the figure on the right is for the simulation using the standard SPH scheme. Density is normalized to 4~g~cm$^{-3}$.}
\label{figure9}
\end{figure}

\begin{figure}
\includegraphics[width=\columnwidth]{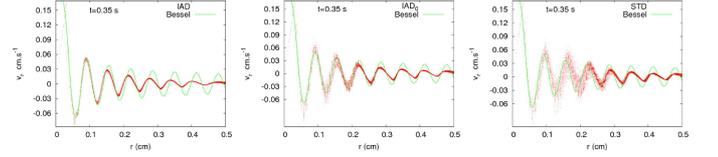}
\caption{Acoustic wave profiles at t=0.35 s calculated using {\sl IAD}, {\sl IAD$_0$} and standard SPH (STD). The waves were generated by the periodic displacement of a circular piston of size 0.035 cm located at the center of the lattice. The continuum line is the analytical solution calculated taking a sound speed of $c_s=1.29$~cm~s$^{-1}$.}
\label{figure10}
\end{figure}
 
\begin{figure}
\includegraphics[width=\columnwidth]{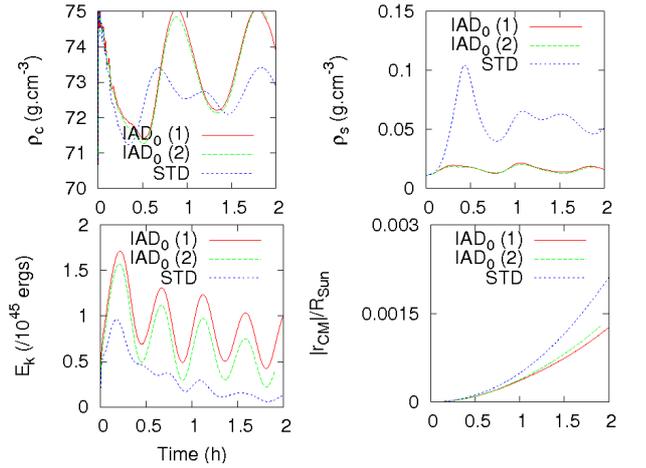}
\caption{Evolution towards stability of a Sun-like star approached by a polytrope. Calculations were carried out using {\sl IAD$_0$}~and the standard, STD, schemes. Labels {\sl IAD$_0$~(1)}~and {\sl IAD$_0$~(2)}~refer to calculations 
with 
the AV parameters set to $\alpha=1,~\beta=2$~and $\alpha=1.5, ~\beta=3$~respectively.} 
\label{figure11}
\end{figure}

\begin{figure}
\includegraphics[width=\columnwidth]{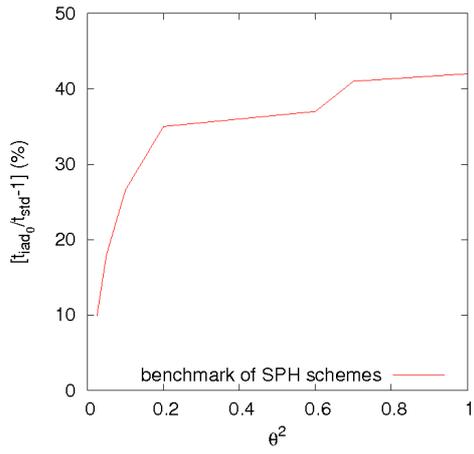}
\caption{Benchmarking of {\sl IAD$_0$} versus STD. Parameter $\theta$~is the tolerance parameter assumed in the multipolar calculation of gravity.}
\label{figure12}
\end{figure}

\clearpage
\begin{table*}[h!]
\centering
\begin{tabular}{@{}llrrrr@{}}
\hline
Test & Scheme & Time (s) & $\vert~\Delta E\vert/E_0$ & $\vert\Delta x_{cm}\vert$/R & $\vert\Delta y_{cm}\vert$/R  \\
\hline
\hline
KH  & IAD$_0$ & 4.0 & $4.0~10^{-8}$ & $-$            & $3~10^{-5}$    \\
KH  & STD     & 4.0 & $1.0~10^{-6}$ & $-$            & $1.3~10^{-4}$  \\
RT  & IAD$_0$ & 5.7 & $1.2~10^{-5}$ & $5~10^{-5}$    & $-$            \\
RT  & STD     & 5.7 & no grow       & no grow        & $-$            \\
WH  & IAD$_0$ & 0.3 & $5.9~10^{-3}$ & $8~10^{-12}$   & $8~10^{-12}$   \\
WH  & STD     & 0.3 & $10^{-2}$     & $5~10^{-10}$   & $5~10^{-10}$   \\
SED & IAD$_0$ & 0.7 & $9.8~10^{-2}$ & $1.3~10^{-10}$ & $1.3~10^{-10}$ \\
SED & STD     & 0.7 & $1.4~10^{-1}$ & $1.5~10^{-10}$ & $1.6~10^{-10}$ \\
\hline
\label{table1}
\end{tabular}
\caption{Conservation properties of the computed models: Kelvin-Helmholtz (KH), Rayleigh-Taylor (RT), the wall heating shock (WH) and Sedov (SED). Conservation of momentum is tracked by normalizing the deviation of the center of mass over the characteristic size of the system R. See Fig.~\ref{figure5} for the temporal evolution of the momentum in the KH case. The method used to calculate the derivatives {\sl IAD$_0$} or  standard, STD, is indicated.}
\end{table*}

\end{document}